# Automatic lesion detection, segmentation and characterization via 3D multiscale morphological sifting in breast MRI

**Hang Min, Darryl McClymont, Shekhar S. Chandra and Stuart Crozier**

School of Information Technology and Electrical Engineering, University of Queensland, Australia.

**Andrew P. Bradley**

Science and Engineering Faculty, Queensland University of Technology, Australia

**Abstract**

Previous studies on computer aided detection/diagnosis (CAD) in 4D breast magnetic resonance imaging (MRI) regard lesion detection, segmentation and characterization as separate tasks, and typically require users to manually select 2D MRI slices or regions of interest as the input. In this work, we present a breast MRI CAD system that can handle 4D multimodal breast MRI data, and integrate lesion detection, segmentation and characterization with no user intervention. The proposed CAD system consists of three major stages: region candidate generation, feature extraction and region candidate classification. Breast lesions are firstly extracted as region candidates using the novel 3D multiscale morphological sifting (MMS). The 3D MMS, which uses linear structuring elements to extract lesion-like patterns, can segment lesions from breast images accurately and efficiently. Analytical features are then extracted from all available 4D multimodal breast MRI sequences, including T1-, T2-weighted and DCE sequences, to represent the signal intensity, texture, morphological and enhancement kinetic characteristics of the region candidates. The region candidates are lastly classified as lesion or normal tissue by the random under-sampling boost (RUSboost), and as malignant or benign lesion by the random forest. Evaluated on a breast MRI dataset which contains a total of 117 cases with 141 biopsy-proven lesions (95 malignant and 46 benign lesions), the proposed system achieves a true positive rate (TPR) of 0.90 at 3.19 false positives per patient (FPP) for lesion detection and a TPR of 0.91 at a FPP of 2.95 for identifying malignant lesions without any user intervention. The average dice similarity index (DSI) is $0.72 \pm 0.15$ for lesion segmentation. Compared with previously proposed lesion detection, detection-segmentation and detection-characterization systems evaluated on the same breast MRI dataset, the proposed CAD system achieves a favourable performance in breast lesion detection and characterization.

## 1. Introduction

Breast magnetic resonance imaging (MRI) is a sensitive imaging tool for breast cancer that is useful in demonstrating the extent of cancer and evaluating breast cancers (Morris & Liberman, 2005). It can provide clear visualization of cancers that are occult on conventional imaging techniques such as mammography and breast ultrasound (DeMartini et al, 2008). A standard breast MRI protocol typically generates 4D multimodal imaging data, containing T1-, T2-weighted sequences and dynamic contrast enhanced (DCE) sequences acquired before and after the injection of contrast agent (Thomassin-Naggara et al, 2012). In clinical practice, factors such as morphology, signal intensity in T1-, T2-weighted images and dynamic contrast enhancement kinetics in the DCE images are often taken into consideration in the detection and characterization of suspicious regions (Kaiser, 2008). However, visual estimation of all these factors on multiple image sequences can be extremely challenging for human interpreters. Therefore, breast MRI computer aided detection/diagnosis (CAD) systems have been developed to facilitate the quantitative analysis and characterization of breast lesions.

A breast MRI CAD system normally consists of the following stages: extracting region candidates that could represent lesions from the breast image; extracting features that reflect the characteristics of the region candidates; and classifying the region candidates as lesion or normal tissue and malignant or benign lesion based on the features. Previously proposed breast MRI CAD systems typically followed two types of workflow: one uses unsupervised region candidate generation approaches with analytically extracted features; the other one uses deep learning (DL) based methods to automatically learn the focus of attention and deep features supervised by the lesion annotations in the training data.

To generate region candidates in an unsupervised analysis, clustering methods, such as mean-shift (McClymont et al, 2014) and k-means (Li et al, 2019), have been used in previous studies. However, performing an exhaustive search on the whole 3D breast volume using these methods can be computationally expensive (Maicas et al, 2017b), and sometimes only preselected 2D slices that contain lesions are used instead of the whole 3D breast image (Amit et al, 2017). User intervention was often required in previous studies for an accurate segmentation of lesions: Dalmış et al. (Dalmış et al, 2016) used manually selected seeds with a 3D region growing approach; Li et al. (Li et al, 2019) stated the segmentation of lesions as region candidates might require manual correction and the final segmentation needed to be approved by radiologists. To discriminate lesions from normal regions, and malignant from benign lesions, analytical features based on the radiologic signs of breast abnormalities in MRI are then extracted on the region candidates and the region candidates are classified into different categories by classifiers. These hand-crafted descriptors normally include intensity, texture, morphology and contrast enhancement kinetic features extracted from multiple MRI sequences (Chen et al, 2004; Dalmış et al, 2016; Gilhuijs et al, 1998; Li et al, 2019).

The DL-based methods can automatically search for the regions of interest and learn the deep high-level features representing the characteristics of lesions without the need for hand-crafted features, however, it is still challenging for DL networks to directly handle 4D multimodal breast MRI data. For instance, the lesion detection and characterization system based on a supervised attention model using ResNet (He et al, 2016) in (Herent et al, 2019) and the u-net (Ronneberger et al, 2015) based lesion segmentation system in (Spuhler et al, 2019) were both developed for 2D pre-selected MRI slices instead of 3D breast volumes. Some other studies did not use all available sequences. For example, Maicas et al. (Maicas et al, 2017b) adopted the deep Q-network (DQN) (Mnih et al, 2015) and ResNet for lesion detection only using the subtraction volume between the first post-contrast and pre-contrast image, neglecting T2-weighted and other DCE images (T1-weighted images were only used for breast extraction). The authors later extended this detection system into a diagnostic system by adding a 3D DenseNet, which still only used the subtraction volume (Maicas et al, 2019).

Previous studies on breast MRI CADs mostly treated lesion detection, segmentation and malignancy classification as separate tasks and may require user intervention. Maicas et al. (Maicas et al, 2017b) and Amit et al. (Amit et al, 2017) only focused on breast lesion detection. Dalmış et al. (Dalmış et al, 2016) only classified the manually selected lesions as malignant or benign. McClymont et al. (McClymont et al, 2014) presented a lesion detection and segmentation system which required each parameter to be tuned sequentially. Maicas et al. (Maicas et al, 2017a) combined a lesion detection DL network (Dhungel et al, 2015) and a segmentation network, where the two networks were tuned separately and the lesion segmentation was only performed on the true positive detections manually confirmed by users. Both Herent et al. (Herent et al, 2019) and Maicas et al. (Maicas et al, 2019) only performed lesion detection and characterization without segmentation, and the method in Herent et al. (Herent et al, 2019) could only handle 2D breast MRI slices. Spuhler et al. (Spuhler et al, 2019) only performed lesion segmentation on 2D MRI slices.

This work aims at integrating lesion detection, segmentation and characterization on 4D multimodal breast MRI with no need for user intervention. Figure 1 shows the diagram of the proposed breast MRI CAD system. A novel efficient region candidate proposal method, 3D multiscale morphological sifting (3D MMS), is introduced in this study. The 3D MMS uses sets of morphological filters with linear structuring elements at multiple scales, which is a 3D extension of the 2D MMS proposed in our previous work (Min et al, 2019). The 3D MMS can segment lesions as region candidates from 3D breast images accurately and efficiently, which enables the CAD to directly evaluate the regions' morphology, intensity, texture and enhancement kinetic characteristics extracted from all available MRI sequences, i.e. T1-, T2-weighted and DCE images. To assess the lesion detection, segmentation and malignancy identification performance, the proposed integrated system is evaluated on training and testing datasets the same as three previously proposed CAD systems which are lesion detection (Maicas et al, 2017b), detection-segmentation (Maicas et al, 2017a) and detection-characterization systems (Maicas et al, 2019). The proposed system achieves a favourable performance in lesion detection and characterization than these previous systems.

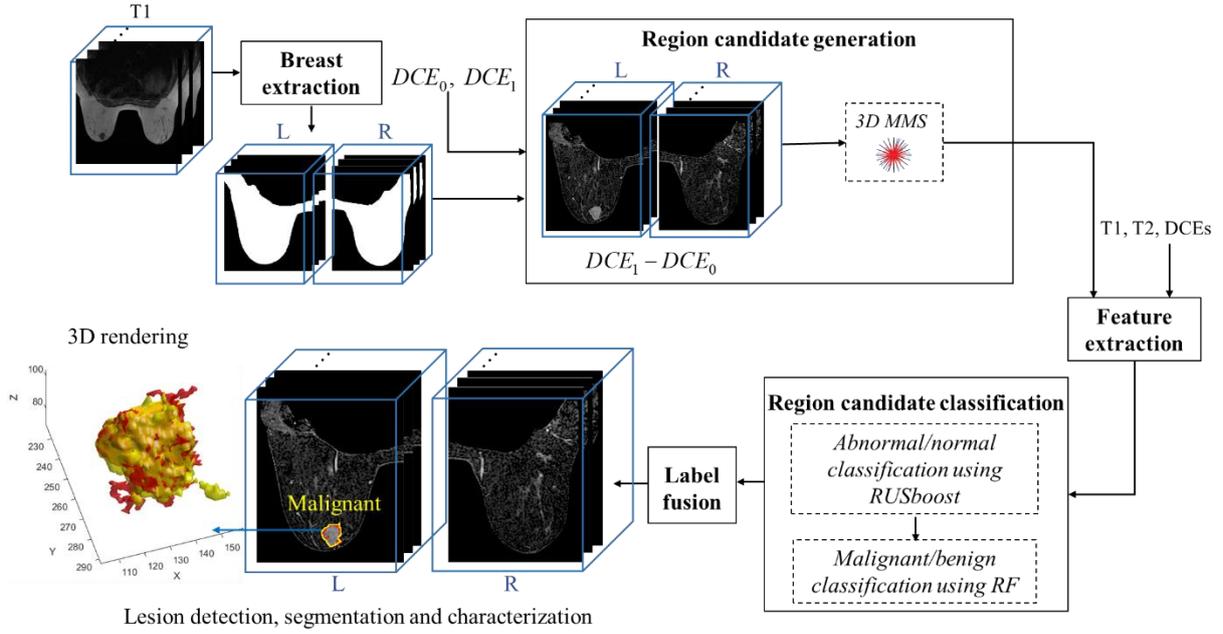

Figure 1. The diagram of the proposed breast MRI CAD system. The 3D MMS (multiscale morphological sifting) extracts region candidates from the subtraction image $DCE_1 - DCE_0$, where $DCE_0$ and $DCE_1$ represent the pre- and first post-contrast image respectively. Analytical features are extracted from region candidates on all available sequences, including $T_1$, $T_2$ and DCE sequences (DCEs). Region candidates are classified as lesion or normal region, and as malignant or benign by the RUSboost (random under-sampling boost) and RF (random forest) respectively. Final segmentation and characterization outputs are generated after the label fusion stage. L and R indicate whether the image represents a left or right breast. The ground truth is outlined in red. The detected/segmented region is identified as a malignant lesion in this case and outlined in yellow.

## 2 Methods

The methodology of the proposed breast MRI CAD system is presented in this section. As shown in Figure 1, the proposed system consists of pre-processing, region candidate generation, feature extraction, region candidate classification and label fusion stages. Firstly, the breast region is segmented from the MRI image in the pre-processing stage. Secondly, region candidates are extracted using 3D MMS. Analytical features are then extracted on these region candidates from all available MRI sequences (T1-, T2-weighted and DCE images). Since the 3D MMS generates many more region candidates representing normal tissues than lesions, the random under-sampling boost (RUSboost) (Seiffert et al, 2010) is used to classify the region candidates while handling class imbalance. A random forest (RF) (Breiman, 2001) classifier is then used to classify the lesions as malignant or benign. Finally, the lesion detection, segmentation and malignancy classification outcomes are generated after the label fusion stage.

### 2.1 Breast MRI Dataset

The breast MRI data (Mcclymont, 2015; McClymont et al, 2014) used in this study was obtained from Queensland X-ray with the approval of the Human Research Ethics Committee at the University of Queensland. This dataset contains 117 cases which were acquired with a 1.5 Tesla GE Signa HDxt scanner. The age of the patients ranges from 24 to 81 year-old and the imaging protocol consists of a T1-weighted, a T2-weighted and up to five DCE sequences. The T1-weighted images were acquired axially without fat suppression with the following parameters: echo time TE = 8 ms, repetition time TR = 480 ms, flip angle of 90° and acquisition matrix = 512 × 512. The T2-weighted sequences were acquired with fat suppression with the following parameters: TE = 46 ms, TR = 6600 ms, flip angle of 90° and an acquisition matrix of 320 × 224. The DCE sequences were acquired as T1-weighted images with fat suppression with the following parameters: TE = 3.4 ms, TR= 6.5 ms, flip angle = 10° and the acquisition matrix ranges from 224 × 224 to 360 × 360. Each DCE sequence was acquired in approximately 90 seconds with a 45 seconds delay between the pre-contrast and the first post-contrast sequence (Mcclymont, 2015). The post-contrast images were acquired after the injection of Gadopentate dimeglumine, 0.2 mmol/kg, administered at a rate of about 2 ml/s using a pressure injector. The DCE image resolution ranges from

$0.7 \times 0.7 \times 1.3\ mm^3$ to $0.55 \times 0.55 \times 1\ mm^3$. The T1-, T2-weighted and the DCE sequences were co-registered to the first post-contrast sequence using the fast non-rigid B-spline registration module in 3D Slicer (www.slicer.org) (Fedorov et al, 2012; Mcclymont, 2015; McClymont et al, 2014). One hundred and forty-one lesions were biopsied under either MRI or ultrasound guidance, with 95 malignant and 46 benign. The diameter range of the lesions is $4 \sim 63 mm$. Table 1 shows the percentage of histological types and subtypes of the lesions in the dataset. There are 71 lesions showing mass-like (ML) enhancement, 35 lesions showing non-mass-like (NML) enhancement, and the rest are not specified. The ground truth was segmented by an experienced radiographer with over 12 years of experience in breast MRI using a region growing tool in Osirix (Rosset et al, 2004).

Table 1. Histological types and subtypes of the lesions in the breast MRI dataset.

| Histological types | Histological subtypes | Percentage (%) |
|---|---|---|
| Malignant | Invasive Ductal Carcinoma (IDC) | 23.4 |
| | Invasive Lobular Carcinoma (ILC) | 7.1 |
| | Ductal Carcinoma in Situ (DCIS) | 17.7 |
| | Others | 16.3 |
| Benign | Fibrocystic changes (FBCY) | 7.1 |
| | Fibroadenoma (FIBD) | 6.4 |
| | Hyperplasia | 3.6 |
| | Others | 18.4 |

*2.2 Pre-processing*

The breast region is extracted by Otsu thresholding (Otsu, 1979) and Hayton's algorithm (M.Hayton, 1998) from T1-weighted images as described in McClymont et al. (McClymont et al, 2014). The breast volume is split into left and right breasts from the middle line of the breast mask, and the proposed method processes one breast at a time, similar to Maicas et al. (Maicas et al, 2019; Maicas et al, 2017a; Maicas et al, 2017b).

*2.3 Region candidate generation via 3D MMS*

The region candidates are extracted from the subtraction image between the first post-contrast DCE image $DCE_1$ and the pre-contrast image $DCE_0$. To generate region candidates, the 3D multiscale morphological sifting (3D MMS) method is proposed. The 3D morphological sifting (MS) is an extension to the 2D MS which uses linear structuring elements to extract lesion-like patterns (Min et al, 2019). The diagram of region candidate generation is shown in Figure 2. To reduce the computational time, region candidates are extracted at 3 different scales (one original scale and two smaller ones). The scaled images are filtered by 3D MS, and then processed by a multilevel intensity thresholding (MLT) (Liao et al, 2001; Otsu, 1979) and a size thresholding.

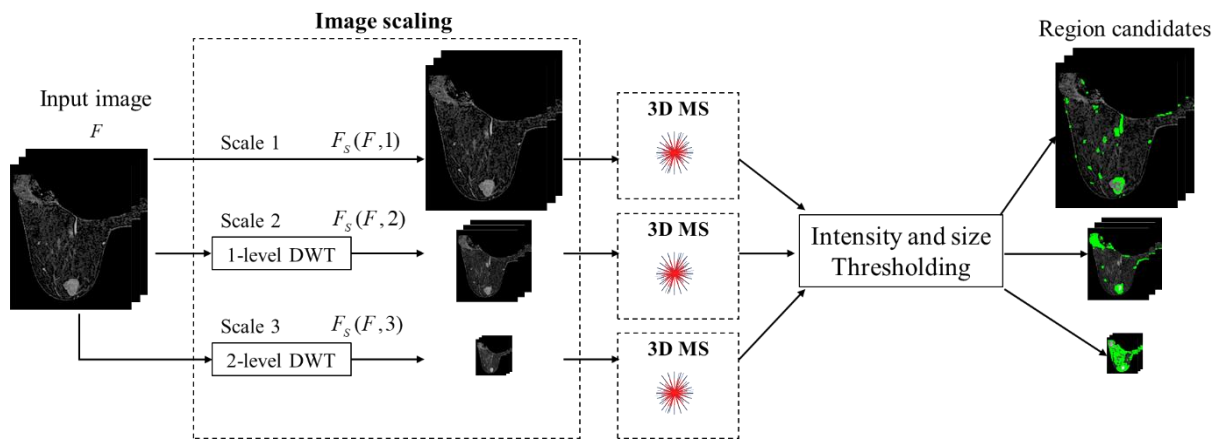

Figure 2. The diagram of region candidate generation on breast MRI images using 3D MMS. The input image F is firstly scaled into 3 resolutions and $F_S$ stands for the scaled image. Then the scaled images are processed by the 3D MS, an intensity and a size thresholding. The region candidates generated are outlined in green.

Three scales, including an original scale, 1/2 and 1/4 of the original scale, are used in 3D MMS as shown in Figure 2. The 3D discrete wavelet transform (DWT) with Daubechies 2 (db2) wavelet is adopted to scale the MRI image and the output from only the low-pass filter is used as the scaled result. If the input MRI image is $F$ and $M$ is the number of scales used, the scaled image by DWT at $scale_m$ ($m = 1, ..., M$) is defined as

$$F_S(F, m) = \begin{cases} F, & \text{if } m = 1 \\ LLL(DWT(F, m-1)), & \text{if } m > 1 \end{cases}, \quad (1)$$

where $LLL$ stands for selecting the approximate image from the low pass decomposition filter and $m - 1$ specifies the level of DWT.

The MS on 2D images (Min et al, 2019) is described as

$$MS(f, ML_1, ML_2) = \sum_{n=0}^{N-1} \{f - [f \circ L(ML_2, \theta(n))]\} \circ L(ML_1, \theta(n)), \quad (2)$$

where $f$ stands for the input image. $\{L(ML_1, \theta(n)) | n = 0, 1, ..., N-1\}$ and $\{L(ML_2, \theta(n)) | n = 0, 1, ..., N-1\}$ stand for two sets of linear structuring elements (LSEs), where $ML_1$ and $ML_2$ are the magnitudes and $\theta(n)$ is the orientation of the LSEs.

On 3D images, the MS is carried out on axial, sagittal and coronal direction separately as shown in Figure 3. The 3D MS on one scale is defined as

$$f' = MS3D(f, ML_1, ML_2)$$
$$= MS_{Axial}(f, ML_1, ML_2) + MS_{Sagittal}(f, ML_1, ML_2) + MS_{Coronal}(f, ML_1, ML_2), \quad (3)$$

$$MS_{Axial}(f, ML_1, ML_2) = MS(f, ML_1, ML_2), \quad (4)$$

$$MS_{Sagittal}(f, ML_1, ML_2) = P_{1,3,2}\left(MS(P_{1,3,2}(f), ML_1, ML_2)\right), \quad (5)$$

$$MS_{Coronal}(f, ML_1, ML_2) = P_{3,2,1}\left(MS(P_{3,2,1}(f), ML_1, ML_2)\right), \quad (6)$$

where permutation $P_{p1,p2,p3}$ stands for re-arranging the dimensions of the matrix in the order of $[p1, p2, p3]$. The scaled image by the DWT decomposition $F_S$ is used as the input image for 3D MS, i.e. $f = F_S(F, m)$ on $scale_m$. If the volume range of the target to be detected in $mm^3$ is $[V_{min}, V_{max}]$ and the image resolution is $d \times d \times D\ mm^3$ ($d$: axial plane pixel spacing, $D$: slice spacing), the magnitudes for the LSEs $[ML_1, ML_2]$ in pixels are defined as

$$[ML_1, ML_2] = \begin{cases} \left[\frac{ML'_1}{d}, \frac{ML'_2}{d}\right], & \text{for } MS_{Axial} \\ \left[\min\left(\frac{ML'_1}{d}, \frac{ML'_1}{D}\right), \frac{ML'_2}{d}\right], & \text{for } MS_{Coronal} \text{ and } MS_{Sagittal} \end{cases} \quad (7)$$

$$ML'_1 = \left(\frac{6}{\pi} V_{min}\right)^{1/3}, ML'_2 = \frac{1}{2^{(M-1)}} \left(\frac{6}{\pi} V_{max}\right)^{1/3} \quad (8)$$

Here, $[ML'_1, ML'_2]$ is the diameter range of lesions measured in $mm$. If the spacing $d = D$, the $[ML_1, ML_2]$ will be the same between $MS_{Axial}, MS_{Coronal}$ and $MS_{Sagittal}$. However, $d$ is often smaller than $D$ in practice, and this spacing difference needs to be taken into consideration as in Eq. (7). The output from MS3D is then linearly normalized within the breast to 16-bit.

The MLT is then applied to partition the output image $f'$ from 3D MS into region candidates as shown in Figure 3. Given a series of thresholds $\{Th_t | t = 1, ..., T\}$ generated by MLT, $T$ binary masks are generated. The $t^{th}$ binary mask $B(x, y, z, Th_t)$ is defined as

$$B(x, y, z, Th_t) = \begin{cases} 0, & \text{if } f'(x, y, z) < Th_t \\ 1, & \text{if } f'(x, y, z) \geq Th_t \end{cases}, \quad (9)$$

where $(x, y, z)$ specifies the voxel position on the input image $f'$.

After the intensity thresholding, the region candidates are then sieved by their sizes. On $scale_m$, if the volume ($mm^3$) of a region candidate (RC) satisfies

$$\begin{cases} V_{min} \leq Vol(RC) \leq \frac{V_{max}}{2^{3(M-1)}}, & m = 1 \\ \frac{V_{max}}{2^{3(M-m+1)}} \leq Vol(RC) \leq V_{max}, & m > 1 \end{cases} \quad (10)$$

it is kept as a region candidate, otherwise discarded.

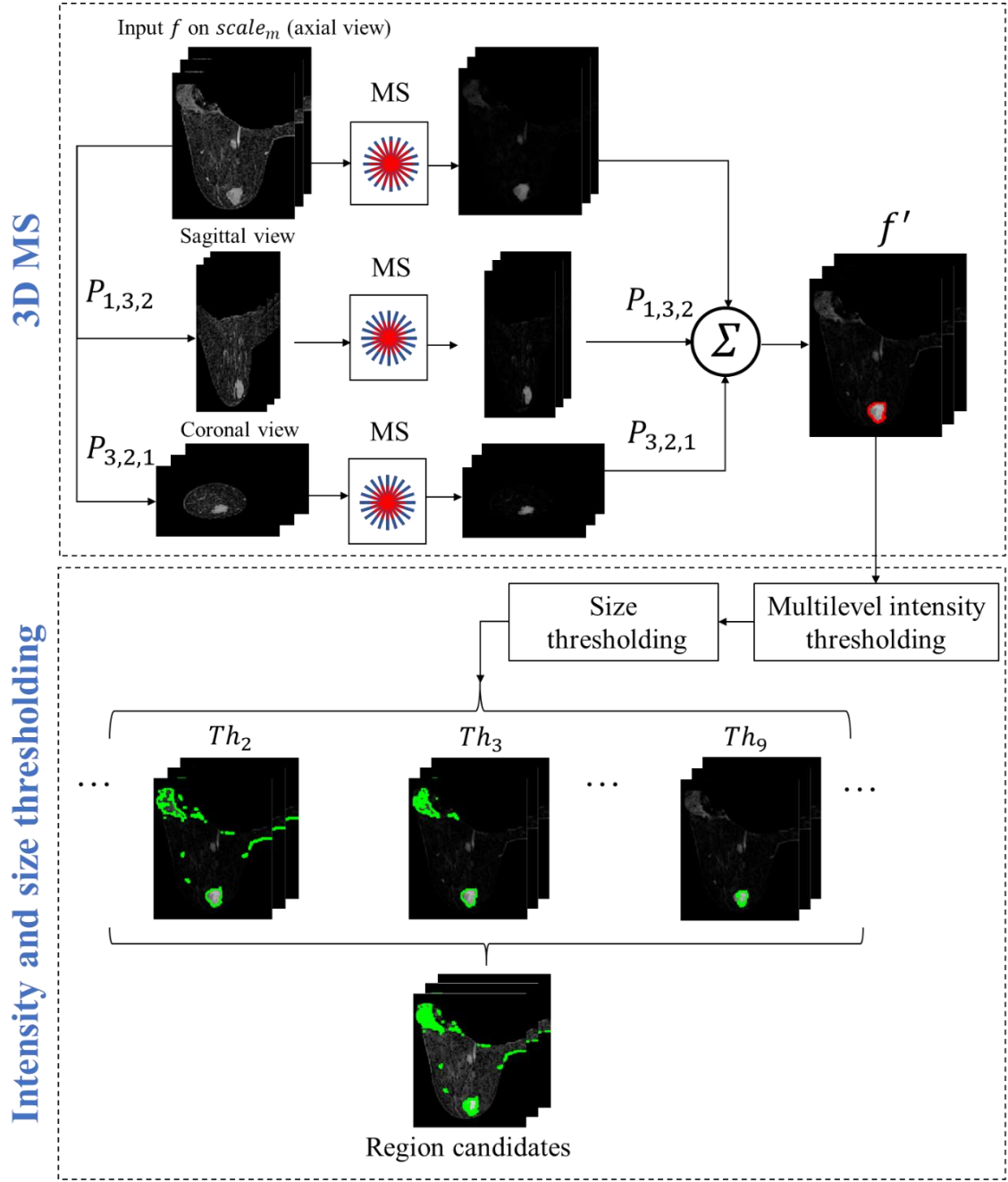

Figure 3. Region candidate generation using 3D MS on a single scale. The input image $f$ is permuted into sagittal and coronal view, and passed into MS. The outputs from MS are then permuted back to axial view and summed up. $P_{p1,p2,p3}$ stands for permuting the dimensions of the matrix in the order of $[p1, p2, p3]$. MS stands for morphological sifting. Region candidates generated are outlined in green, and the lesion ground truth is outlined in red. $Th_t$ stands for the threshold generated by multilevel thresholding.

The 3D MMS only consists of 3 parameters, the number of scales $M$, number of LSEs $N$ and number of intensity thresholds $T$, which are relatively simple to set. $M$ is set as 3, which accommodates the size variance of the lesions well with a reasonable computational complexity similar to (Maicas et al, 2019; Maicas et al, 2017b). The number of LSEs $N$ effects the smoothness of the MMS filtered image. As $N$ increases, the smoother the processed image becomes at the cost of an increased computational complexity. $N = 10$ is a balanced value that generates relatively smooth processed image at a reasonable computational cost (Min et al, 2019). As to the number of thresholds $T$, the larger $T$ is, the more region candidates will be generated and the more likely for the algorithm to capture the accurate outlines of the lesions, but with a higher computational complexity. When $T =$

16, there are enough levels of threshold to segment the lesions in the training set accurately, and increase in $T$ does not result in significant changes in segmentation (Min et al, 2019).

*2.4 Feature extraction*

Four categories of features including intensity, texture, morphology and contrast enhancement kinetics are used in this work as shown in Table 2. They are radiomic features proposed in previous studies to quantize the radiologic signs of lesions on different breast MRI sequences. These features are calculated for the region candidates on the T1-, T2-weighted, DCE images or the 3D binary mask of region candidates as specified in Table 2. The image sequences need to be scaled to match the region candidates extracted at different scales. The intensity and texture features are adopted, since lesions and normal breast tissues can display different intensities and patterns in different sequences (Kaiser, 2008). The T1-, T2-weighted and pre-contrast DCE image ($DCE_0$) are normalized to the mean intensity of the fat tissue (Mcclymont, 2015) when intensity features are calculated. The fat tissue mask is extracted by Otsu thresholding on the T1-weighted image (Mcclymont, 2015). The margin in margin sharpness is a $3mm$ shell with $1mm$ inside and $2mm$ outside the region candidate (Mcclymont, 2015). Morphological features analyse the shape and margin of a region candidate, which are important aspects in the characterization of breast lesions (Kaiser, 2008). The equivalent spherical diameter is measured in $mm$. The kinetic features, such as enhancement kinetics and parametric model, analyse the average signal intensity changes in a region overtime (Chen et al, 2004; Mcclymont, 2015). Blooming sign and peripheral uptake are designed to analyse the internal and margin enhancement patterns (Agliozzo et al, 2012; Mcclymont, 2015).

Table 2. Intensity, texture, morphological and kinetic features extracted from T1-, T2-weighted, DCE images and the 3D binary mask of region candidates. $DCE_0$ and $DCE_1$ stand for the pre-contrast and first post-contrast image respectively. DCEs stands for the DCE images.

| Feature type | Features | Modality |
|---|---|---|
| Intensity | Mean (Bhooshan et al, 2011) | T1, T2, $DCE_0$ |
| | Standard deviation (Sutton et al, 2015) | |
| | Kurtosis (Pearson, 1905; Sutton et al, 2015) | T1, T2 |
| | Skewness (Pearson, 1895; Sutton et al, 2015) | |
| | 20th percentile (Van Aalst et al, 2008) | T2 |
| | 90th percentile (Van Aalst et al, 2008) | |
| | Presence of edema (92nd percentile in 2 $mm$ shell, and 98th percentile in 10, 20 $mm$ shell) (Van Aalst et al, 2008) | |
| Texture | Haralick features (Coelho, 2012; Haralick & Shanmugam, 1973) | T2, $DCE_1$, $DCE_1 - DCE_0$ |
| | Margin sharpness (Gilhuijs et al, 1998) | T2, DCEs |
| | Radial gradient index (Gilhuijs et al, 1998) | |
| Morphology | Equivalent spherical diameter (Gilhuijs et al, 1998) | Binary region |
| | Extent (Sutton et al, 2015) | |
| | Solidity (Sutton et al, 2015) | |
| | Irregularity (Gilhuijs et al, 1998) | |
| | Fat fraction (Mcclymont, 2015) | |
| Kinetics | Enhancement kinetics (Chen et al, 2004) | DCEs |
| | Enhancement-variance dynamics (Chen et al, 2004) | |
| | Enhancement parametric model (Mcclymont, 2015) | |
| | Blooming sign (Mcclymont, 2015) | |
| | Peripheral uptake (Agliozzo et al, 2012) | |

*2.5 Region candidate classification*

In this stage, the region candidates represented by their feature vectors are classified as lesion or normal tissue by RUSboost, and malignant or benign lesion by RF. Both RUSboost and RF have a built-in feature selection mechanism which selects the important features that can minimize the classification error during training (Breiman, 2001; Seiffert et al, 2010).

The RUSboost, an ensemble learning method that can handle severe class imbalance (Seiffert et al, 2010), is adopted to classify the regions as lesion or normal tissue, since there are $1.6 \times 10^4$ negative (normal) region candidates and only 76 positive (lesion) ones extracted from the training set. Here, the number of positive region candidates is slightly larger than the number of lesions in the training set (72) since a few lesions appear segmental and contain more than one connected region. To be labelled as positive, a region candidate in the training set has to satisfy: 1) it has the highest Dice similarity index (DSI) (Dice, 1945) to the ground truth among all the candidates, 2) the DSI is no less than 0.6 (Maicas et al, 2017b; Min et al, 2019). If a region candidate has a DSI to the lesion annotation smaller than 0.2, it is labelled as negative. The rest of the region candidates are neutral and not used in the training of RUSboost. Decision trees are used as the base classifier for RUSboost. The learning rate of RUSboost is set as 0.1 and the maximum number of decision splits is set as the number of training samples (Min et al, 2019). The number of trees (NoT) in RUSboost is decided by performing a 5-fold cross-validation on the training set and calculating the classification loss. The classification loss is defined as the mean square error (MSE) between the predicated value of the sample and its true label. The loss decreases as the NoT increases and remains constant after the NoT reaches 2000. Therefore, the number of trees in RUSboost is set as 2000 and the RUSboost is trained on all training samples with this setting.

A RF (Breiman, 2001; Jaiantilal, 2014) is then used to further identify the malignancies among the detected lesions. The RF is trained on the 76 region candidates representing lesions (53 malignant and 23 benign regions) in the training set. A grid search with $n_{tree}$ in the range of [100,1000] and $m_{try}$ in the range of $[0.5\sqrt{NF}, 2\sqrt{NF}]$ is carried out, where $n_{tree}$ stands for the number of trees, $m_{try}$ stands for the number of features tried at each split and $NF$ stands for the number of features. The RF model with the lowest MSE is chosen as the final classifier.

*2.6 Label fusion*

After the region candidates are passed through the classifiers, there can be multiple detected regions overlapping with each other. To generate a final segmentation contour, the overlapping regions are fused by selecting the region with the highest probability to be positive as shown in Figure 4. Lastly, the detected region candidates are upscaled to the original image resolution through wavelet reconstruction.

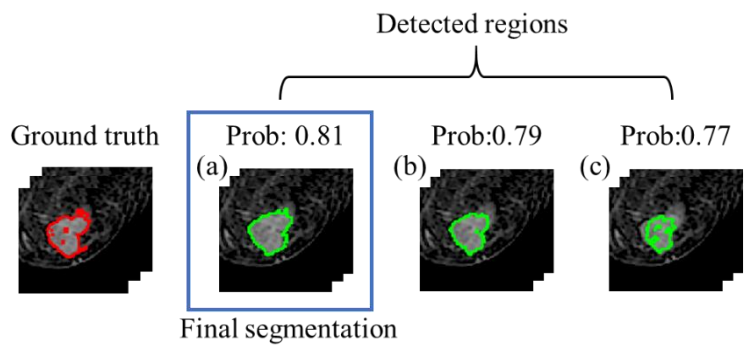

Figure 4. Label fusion for breast MRI CAD. Region (a), (b) and (c) are the regions detected at the lesion area. The ground truth is outlined in red, and the detected regions are outlined in green. Prob stands for the region's probability to be positive.

*2.7 Evaluation methods and experiment settings*

The same training and testing data partition as Maicas et al. (Maicas et al, 2019; Maicas et al, 2017a; Maicas et al, 2017b) is used to ensure a fair comparison, with 58 cases in the training set and 59 cases in the testing set. A total of 68% (49/72) and 67% (46/69) of the lesions are malignant in the training and testing sets respectively. True positive rate (TPR), false positive per patient (FPP), the receiver operating characteristic (ROC), area under the ROC curve (AUC) and the free response operating characteristic (FROC) curve are used to measure the lesion detection and malignancy identification performance. The DSI is used to measure the lesion segmentation performance. For lesion detection and malignancy identification, the target is regarded as detected if the DSI between a detected region and the ground truth is at least 0.2 ($DSI \geq 0.2$), similar to Maicas et al. (Maicas et al, 2019; Maicas et al, 2017b). All experiments are carried out on a Dell desktop with Intel Core i7-4790CPU@3.60 GHz, 16 GB RAM.

# 3. Results

## 3.1 Region candidate generation performance

The 3D MMS generates approximately $274 \pm 104$ region candidates per breast on average. To evaluate how accurately the 3D MMS can extract lesions as region candidates, the accuracy of region candidate generation (ARCG) is measured by the average DSI between the ground truth and the region candidates that have the highest DSI to the ground truth on the whole dataset. The ARCG for 3D MMS in region candidate generation is $0.81 \pm 0.11$ and no lesion is missed in this stage. Each breast takes approximately 17 seconds on average to process. Table 3 shows the region candidate generation performance comparison between the proposed 3D MMS and three exhaustive search clustering methods used in previous studies, which are k-means (Moftah et al, 2014), fuzzy C-means (Chen et al, 2006), and SLIC superpixel (Yu et al, 2015) clustering. The number of clusters is set the same as the average number of region candidates generated per breast by 3D MMS. Figure 5 shows the region candidate generation examples for several ML and NML lesions in various shapes and sizes. The NML lesions are often more challenging to separate from the background than the ML ones since they can appear in highly irregular shapes and even appear to be segmental. Only region candidates with the highest DSI to the ground truth are plotted on Figure 5. The ARCG for ML lesions is $0.86 \pm 0.08$ and $0.73 \pm 0.11$ for NML lesions.

Table 3. Region candidate generation performance comparison between the proposed 3D MMS and several exhaustive search methods.

| Methods | ARCG | Average runtime (per breast) |
|---|---|---|
| K-means (Moftah et al, 2014) | $0.49 \pm 0.21$ | 51s |
| Fuzzy C-means (Chen et al, 2006) | $0.43 \pm 0.28$ | 892s |
| SLIC superpixel (Yu et al, 2015) | $0.14 \pm 0.16$ | **7s** |
| **3D MMS** | $\mathbf{0.81 \pm 0.11}$ | 17s |

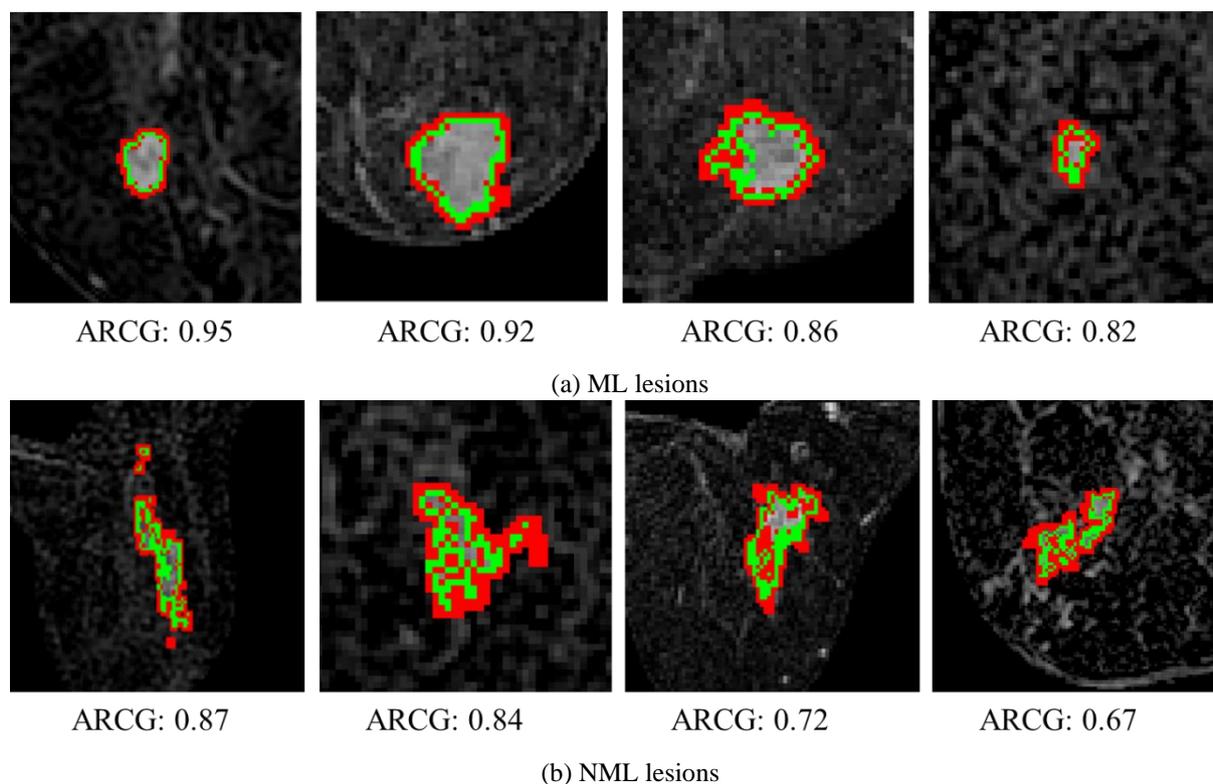

ARCG: 0.95　　ARCG: 0.92　　ARCG: 0.86　　ARCG: 0.82

(a) ML lesions

ARCG: 0.87　　ARCG: 0.84　　ARCG: 0.72　　ARCG: 0.67

(b) NML lesions

Figure 5. Examples of region candidates generated for ML (mass-like) and NML (non-mass-like) lesions in various shapes and sizes. The red lines represent the ground truth, while the green lines represent the region candidates generated at the lesion area. Only region candidates with the highest DSI to the ground truth are plotted. One slice from each lesion is selected as an example, while the ARCG is calculated on 3D regions.

## 3.2 Lesion detection, segmentation and characterization performance

The ROC and FROC curves for lesion detection are shown in Figure 6 (a) and (b). The AUC for lesion detection is 0.99. The system can achieve a TPR of 0.90 at 3.19 FPP as marked with a red dot on Figure 6 (b). The median and average DSI is 0.74 and $0.72 \pm 0.15$ respectively. The malignancy identification is carried out at the operating point where TPR = 0.90 and FPP = 3.19 in lesion detection. The ROC and FROC curve for malignancy identification are shown in Figure 6 (c) and (d). The AUC for malignancy identification is 0.70. The system can achieve a TPR of 0.91 at 2.95 FPP for malignancy identification as marked with a red dot on Figure 6 (d). Figure 7 shows several detection, segmentation and malignancy identification examples of lesions in various shapes and sizes, where the malignancies are labelled in yellow and benign lesions are labelled in cyan. The segmentation of lesions is shown in both 2D slices and 3D rendering. Table 4 compares the lesion detection, segmentation and malignancy identification performance of the proposed system with a lesion detection system (Maicas et al, 2017b), a detection-segmentation system (Maicas et al, 2017a) and a detection-characterization system (Maicas et al, 2019) on the same breast MRI dataset. It is worth noting that Maicas et al. (Maicas et al, 2017a) used a different evaluation criterion from other studies in Table 4, where a lesion is regarded as detected when the detected region has a $DSI \geq 0.4$ to the ground truth. Under this criterion, the proposed method yields a TPR of 0.86, an FPP of 3.20 and an average DSI of $0.74 \pm 0.12$.

Table 4. Lesion detection, segmentation and malignancy identification (characterization) performance comparison with previously published methods evaluated on the same breast MRI dataset.

| Methods | Lesion detection-segmentation | | | Malignancy identification | | User intervention |
|---|---|---|---|---|---|---|
| | TPR | FPP | Avg. DSI | TPR | FPP | |
| **The proposed integrated CAD** | **0.90** | **3.19** | $0.72 \pm 0.15$ | **0.91** | **2.95** | No |
| Detection CAD (Maicas et al, 2017b) | 0.80 | 3.2 | -- | -- | -- | No |
| Detection-segmentation CAD (Maicas et al, 2017a) | 0.85 | 3.66 | $\mathbf{0.77 \pm 0.13}$ | -- | -- | Yes |
| Detection-characterization CAD (Maicas et al, 2019) | 0.80 | 3.2 | -- | 0.80 | 2.9 | No |

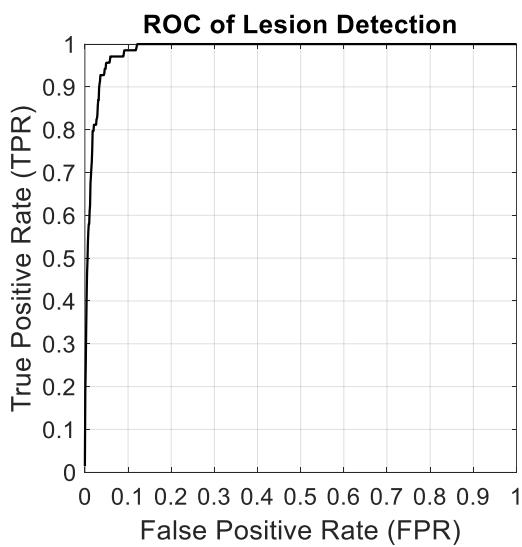
(a)

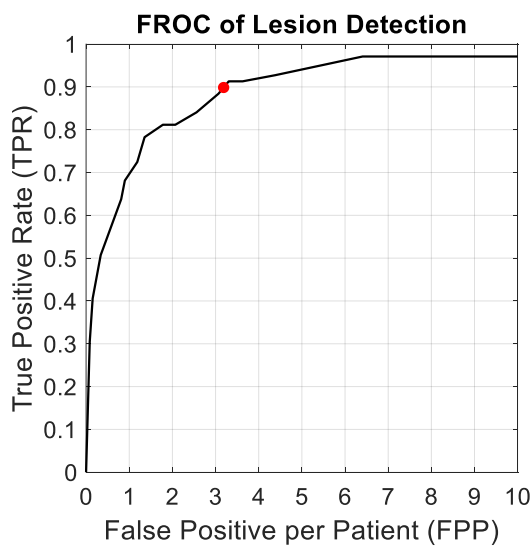
(b)

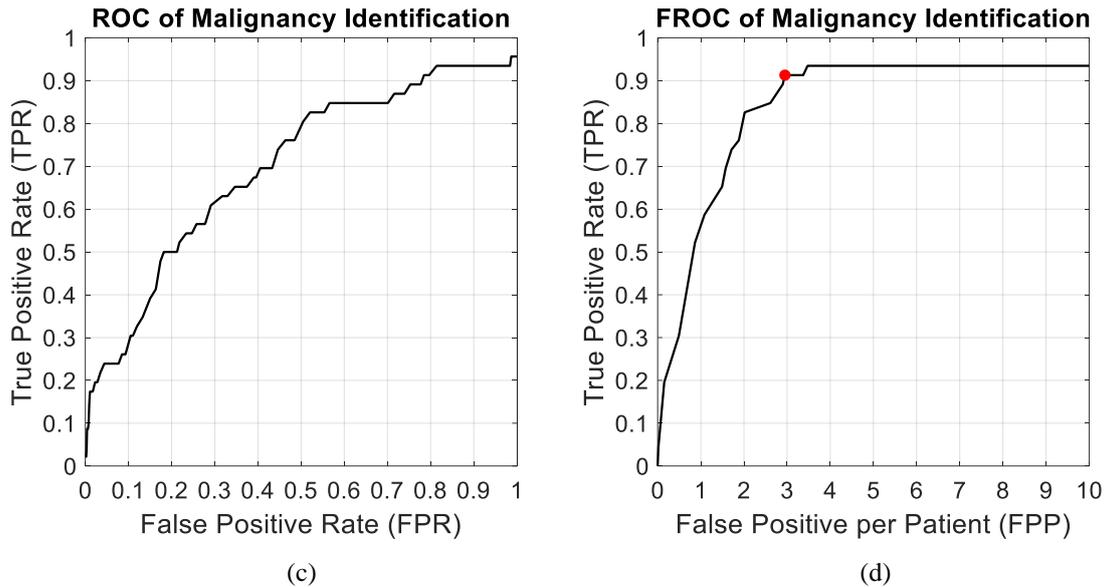

Figure 6. (a) and (b) are the ROC and FROC curves for lesion detection. The red dot marks where $TPR = 0.90$ at $FPP = 3.19$. (c) and (d) are the ROC and FROC curves for malignancy identification. The red dot marks where $TPR = 0.91$ and $FPP = 2.95$.

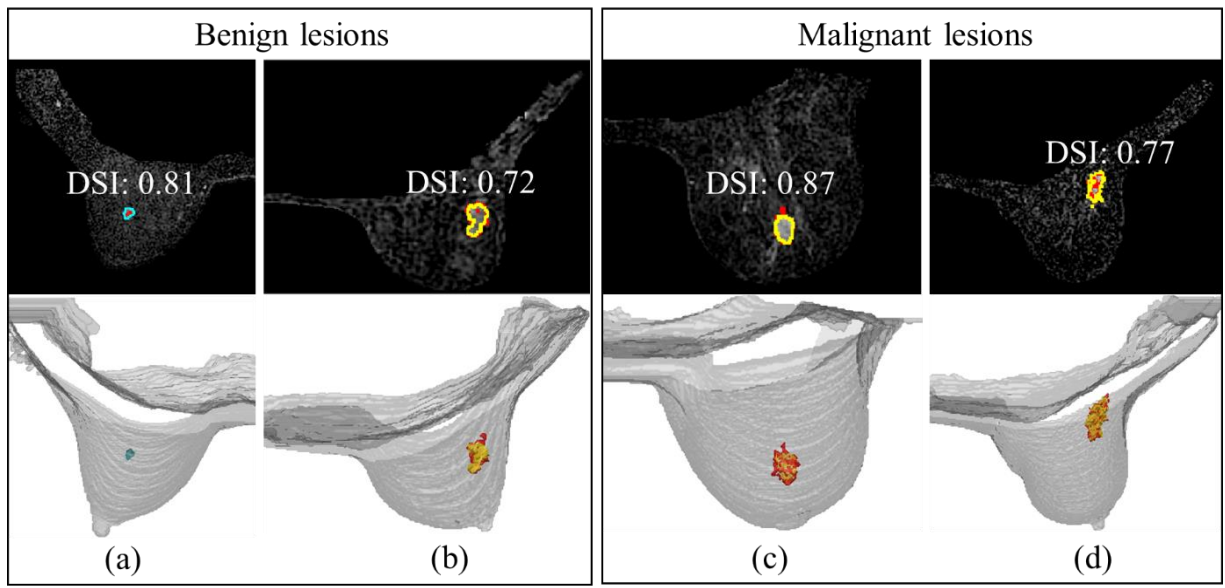

Figure 7. The lesion segmentation and characterization examples. The ground truth of the lesion is marked in red. The detected malignant regions are marked in yellow and the benign ones are marked in cyan. The lesions in (a) and (b) are benign lesions, and the detected region in (b) is a false positive malignant detection.

## 4. Discussion

In this work, an integrated breast MRI CAD system for simultaneous lesion detection, segmentation and characterization is presented. This system can directly process multimodal 4D breast MRI data instead of only processing manually selected 2D MRI slices as in many previous studies (Amit et al, 2017; Herent et al, 2019; Spuhler et al, 2019). This system detects and segments the lesions at the same time, without requiring users to reject the FP detections as in Maicas et al. (Maicas et al, 2017a). Unlike many diagnosis systems which only differentiate malignant and benign lesions without considering the FP detections or use manually selected regions of interest (Dalmış et al, 2016; Lu et al, 2017; Rasti et al, 2017), the proposed method directly identifies the malignant lesions together with the automatic detection and segmentation of lesions.

The proposed system utilizes a novel 3D region candidate generation method, 3D MMS. This method can accurately segment lesions as region candidates at a relatively low computational cost. The 3D MMS extracts region candidates on the original scale and two smaller scales. Using the original scale allows the algorithm to preserve the visibility of small lesions, while using the two smaller scales allows the algorithm to speed up the process for larger lesions. The db2 DWT is adopted to scale the images since it achieves a higher ARCG when paired with MS compared to other image scaling methods such as linear and cubic sampling.

It can be observed from Table 3 that the proposed 3D MMS can extract lesions as region candidates more accurately than the other methods (Chen et al, 2006; Moftah et al, 2014; Yu et al, 2015). SLIC superpixel has a lower time complexity than the 3D MMS, however, it cannot accurately extract lesions as region candidates from the whole breast image. We also explored the mean-shift clustering which was adopted in McClymont et al. (McClymont et al, 2014), however, it can take over 3000 seconds to process one breast image. The extremely high time complexity makes it impractical to apply this method on the whole dataset. Compared with DL-based region candidate proposal networks (Herent et al, 2019; Maicas et al, 2019; Maicas et al, 2017b), the 3D MMS is an unsupervised approach which does not require training. It can capture the outline of the lesions instead of generating bounding box region candidates as in DL-based methods (Herent et al, 2019; Maicas et al, 2019; Maicas et al, 2017b).

In this work, all available MRI sequences, T1-weighted (non-fat suppression), T2-weighted and DCE volumes, are involved in the analysis of region candidates. Many previous studies on breast MRI CAD only used DCE sequences (Amit et al, 2017; Dalmış et al, 2016; Li et al, 2019; Maicas et al, 2019; Maicas et al, 2017b; Rasti et al, 2017; Yu et al, 2015), however, in a standard breast MRI protocol, the T1- and T2-weighted imaging are routinely performed (Mann et al, 2019). In our case, the inclusion of features from T1- and T2-weighted sequences increases the AUC of the lesion detection ROC from 0.97 using only the DCE sequences to 0.99. The AUC of the malignancy identification ROC improves from 0.68 using only the DCE sequences to 0.70 using features from all sequences. The inclusion of these sequences shows the ability of providing valuable complementary information to facilitate both lesion detection and characterization (Bhooshan et al, 2011; Mann et al, 2019; Van Aalst et al, 2008). In the region candidate classification stage, the RUSboost is adopted to distinguish lesions from normal tissues given the fact that the number of negative samples (normal tissues) is two orders of magnitude larger than the one of positive samples (lesions). If the RUSboost is replaced with a RF which is not specially designed for handling high class imbalance, the AUC of ROC drops to 0.96 compared with 0.99 using RUSboost. However, the RF can achieve a higher AUC (0.70) than RUSboost (AUC = 0.68) for malignancy/benign classification, when the positive (malignant) and negative (benign) classes are much more balanced (with a malignant to benign ratio of approximately 2.3:1). The RF also outperforms support vector machine (AUC = 0.68) and neural network (AUC = 0.66) in malignancy classification on our dataset.

As shown in Table 4, our method yields a much higher TPR at a similar FPP compared to the lesion detection system (Maicas et al, 2017b) when evaluated on the same dataset. This is expected since Maicas et al. (Maicas et al, 2017b) only used the subtraction between the first post-contrast and the pre-contrast sequences without involving the T2-weighted sequence or other post-contrast sequences for kinetic enhancement analysis. The lesion detection-segmentation system proposed by Maicas et al. (Maicas et al, 2017a) consists of separate DL networks for lesion detection and segmentation, involving multiple convolutional neural networks (CNNs), a region candidate merging stage, a random forest and a level-set based algorithm, while the proposed method has a much simpler workflow. Performance-wise, the proposed CAD system achieves a favourable lesion detection performance and identifies malignancies additionally but at a lower DSI for lesion segmentation, since Maicas et al. (Maicas et al, 2017a) only performed segmentation on the TP detections after the FPs had been manually rejected by the users. The lesion detection-characterization system (Maicas et al, 2019) extended the lesion detection system in Maicas et al. (Maicas et al, 2017b) by adding a 3D DenseNet to further identify malignancies among the detected regions. The proposed method outperforms Maicas et al. (Maicas et al, 2019) in both lesion detection and malignancy identification, while providing segmentation of the lesions additionally. It is worth noting that both the proposed system and Maicas et al. (Maicas et al, 2019) distinguish malignant lesions from benign lesions and FP detections generated by the lesion detection stage, which is different from many previous diagnosis systems (Dalmış et al, 2016; Li et al, 2019) that only differentiated between malignant and benign lesions.

There are, however, limitations in this work. The size of the MRI dataset used for evaluation is limited and all the cases contain at least one lesion, which is a common problem for many studies on breast MRI CAD (Maicas et al, 2017a; Maicas et al, 2017b; McClymont et al, 2014). Apart from Maicas et al. (Maicas et al, 2019; Maicas et al, 2017a; Maicas et al, 2017b) which used the same breast MRI dataset, many recent studies used a similar size dataset for evaluation as this work, such as Rasti et al. (Rasti et al, 2017) (112 cases) and Li et al. (Li et al, 2019) (112 cases). In future work, we would like to acquire more MRI examinations, with and without lesions, to evaluate the proposed system. The system currently processes each breast individually. Bilateral features that evaluate bilateral symmetry of both breasts could be included in the future (Li et al, 2019). Image registration was still not included as part of the proposed CAD system, the same as many previous studies (Bhooshan et al, 2011; Maicas et al, 2019; Maicas et al, 2017b; Spuhler et al, 2019). The T1- and T2-weighted sequences originally co-registered to the first post-contrast DCE sequence with 3D Slicer by McClymont et al. (McClymont et al, 2014) were not well aligned in around 20% of the cases, and they were corrected with the Elastix library using rigid and affine registration (Marstal et al, 2016). In future work, we would like to include a more robust cross-modality registration method (Rivest-Hénault et al, 2015) into the pipeline. The proposed system achieves a relatively moderate AUC for malignancy identification in comparison to the one for lesion detection. In addition to manually derived features, we would like to investigate possible ways of incorporating deep leaning to learn more discriminative features directly from breast MRI sequences to improve the malignancy classification performance in the future.

## 5. Conclusion

In this paper, we have proposed an integrated breast MRI CAD system that detects, segments lesions and identifies malignancies simultaneously without user intervention. This work introduced a novel region candidate generation algorithm, the 3D MMS, which aims at segmenting lesions from the whole 3D breast volume efficiently and accurately and enables the CAD system to directly evaluate the region candidates' radiologic characteristics on all available breast MRI sequences. The proposed method shows favourable performance in breast MRI lesion detection and malignancy identification compared to the state-of-the-art methods evaluated on the same dataset.

## Acknowledgments

We would like to thank Dr Gabriel Maicas for providing the breast MRI validation sets used in this work.

## References

Agliozzo, S., De Luca, M., Bracco, C., Vignati, A., Giannini, V., Martincich, L., Carbonaro, L., Bert, A., Sardanelli, F. & Regge, D. (2012) Computer‐aided diagnosis for dynamic contrast‐enhanced breast MRI of mass‐like lesions using a multiparametric model combining a selection of morphological, kinetic, and spatiotemporal features. *Medical physics*, 39(4), 1704-1715.

Amit, G., Hadad, O., Alpert, S., Tlusty, T., Gur, Y., Ben-Ari, R. & Hashoul, S. (2017) Hybrid mass detection in breast MRI combining unsupervised saliency analysis and deep learning, *International Conference on Medical Image Computing and Computer-Assisted Intervention*. Springer.

Bhooshan, N., Giger, M., Lan, L., Li, H., Marquez, A., Shimauchi, A. & Newstead, G. M. (2011) Combined use of T2‐weighted MRI and T1‐weighted dynamic contrast‐enhanced MRI in the automated analysis of breast lesions. *Magnetic resonance in medicine*, 66(2), 555-564.

Breiman, L. (2001) Random forests. *Machine learning*, 45(1), 5-32.

Chen, W., Giger, M. L. & Bick, U. (2006) A fuzzy c-means (fcm)-based approach for computerized segmentation of breast lesions in dynamic contrast-enhanced mr images1. *Academic radiology*, 13(1), 63-72.

Chen, W., Giger, M. L., Lan, L. & Bick, U. (2004) Computerized interpretation of breast MRI: Investigation of enhancement‐variance dynamics. *Medical physics*, 31(5), 1076-1082.

Coelho, L. P. (2012) Mahotas: Open source software for scriptable computer vision. *arXiv preprint arXiv:1211.4907*.

Dalmış, M. U., Gubern‐Mérida, A., Vreemann, S., Karssemeijer, N., Mann, R. & Platel, B. (2016) A computer‐aided diagnosis system for breast DCE‐MRI at high spatiotemporal resolution. *Medical physics*, 43(1), 84-94.

DeMartini, W., Lehman, C. & Partridge, S. (2008) Breast MRI for cancer detection and characterization: a review of evidence-based clinical applications. *Academic radiology*, 15(4), 408-416.

Dhungel, N., Carneiro, G. & Bradley, A. P. (2015) Automated Mass Detection in Mammograms using Cascaded Deep Learning and Random Forests, *2015 International Conference on Digital Image Computing: Techniques and Applications (DICTA)*. IEEE.

Dice, L. R. (1945) Measures of the amount of ecologic association between species. *Ecology*, 26(3), 297-302.

Fedorov, A., Beichel, R., Kalpathy-Cramer, J., Finet, J., Fillion-Robin, J.-C., Pujol, S., Bauer, C., Jennings, D., Fennessy, F. & Sonka, M. (2012) 3D Slicer as an image computing platform for the Quantitative Imaging Network. *Magnetic resonance imaging*, 30(9), 1323-1341.

Gilhuijs, K. G., Giger, M. L. & Bick, U. (1998) Computerized analysis of breast lesions in three dimensions using dynamic magnetic‐resonance imaging. *Medical physics*, 25(9), 1647-1654.

Haralick, R. M. & Shanmugam, K. (1973) Textural features for image classification. *IEEE Transactions on systems, man, and cybernetics*(6), 610-621.

He, K., Zhang, X., Ren, S. & Sun, J. (2016) Deep residual learning for image recognition, *Proceedings of the IEEE conference on computer vision and pattern recognition*.

Herent, P., Schmauch, B., Jehanno, P., Dehaene, O., Saillard, C., Balleyguier, C., Arfi-Rouche, J. & Jégou, S. (2019) Detection and characterization of MRI breast lesions using deep learning. *Diagnostic and interventional imaging*, 100(4), 219-225.

Jaiantilal, A. (2014) *Random forest (regression, classification and clustering) implementation for Matlab (and standalone)*, 2014. Available online: https://github.com/jrderuiter/randomforest-matlab [Accessed.

Kaiser, W. A. (2008) *Signs in MR-mammography*Springer.

Li, H., Sun, H., Liu, S., Zhang, W., Arukalam, F. M., Ma, H. & Qian, W. (2019) Assessing the performance of benign and malignant breast lesion classification with bilateral TIC differentiation and other effective features in DCE‐MRI. *Journal of Magnetic Resonance Imaging*.

Liao, P.-S., Chen, T.-S. & Chung, P.-C. (2001) A fast algorithm for multilevel thresholding. *J. Inf. Sci. Eng.*, 17(5), 713-727.

Lu, W., Li, Z. & Chu, J. (2017) A novel computer-aided diagnosis system for breast MRI based on feature selection and ensemble learning. *Computers in biology and medicine*, 83, 157-165.

M.Hayton, P. (1998) *Analysis of contrast-enhanced breast MRI*. PhD University of Oxford, U.K.

Maicas, G., Bradley, A. P., Nascimento, J. C., Reid, I. & Carneiro, G. (2019) Pre and post-hoc diagnosis and interpretation of malignancy from breast DCE-MRI. *Medical Image Analysis*, 58, 101562.

Maicas, G., Carneiro, G. & Bradley, A. P. (2017a) Globally optimal breast mass segmentation from DCE-MRI using deep semantic segmentation as shape prior, *Biomedical Imaging (ISBI 2017), 2017 IEEE 14th International Symposium on*. IEEE.

Maicas, G., Carneiro, G., Bradley, A. P., Nascimento, J. C. & Reid, I. (2017b) Deep Reinforcement Learning for Active Breast Lesion Detection from DCE-MRI, *International Conference on Medical Image Computing and Computer-Assisted Intervention*. Springer.

Mann, R. M., Cho, N. & Moy, L. (2019) Breast MRI: State of the art. *Radiology*, 292(3), 520-536.

Marstal, K., Berendsen, F., Staring, M. & Klein, S. (2016) SimpleElastix: A user-friendly, multi-lingual library for medical image registration, *Proceedings of the IEEE Conference on Computer Vision and Pattern Recognition Workshops*.

Mcclymont, D. (2015) *Computer assisted detection and characterisation of breast cancer in MRI*. PhD University of Queensland.

McClymont, D., Mehnert, A., Trakic, A., Kennedy, D. & Crozier, S. (2014) Fully automatic lesion segmentation in breast MRI using mean‐shift and graph‐cuts on a region adjacency graph. *Journal of Magnetic Resonance Imaging*, 39(4), 795-804.

Min, H., Chandra, S. S., Crozier, S. & Bradley, A. P. (2019) Multi-scale sifting for mammographic mass detection and segmentation. *Biomedical Physics & Engineering Express*, 5(2).

Mnih, V., Kavukcuoglu, K., Silver, D., Rusu, A. A., Veness, J., Bellemare, M. G., Graves, A., Riedmiller, M., Fidjeland, A. K. & Ostrovski, G. (2015) Human-level control through deep reinforcement learning. *Nature*, 518(7540), 529.

Moftah, H. M., Azar, A. T., Al-Shammari, E. T., Ghali, N. I., Hassanien, A. E. & Shoman, M. (2014) Adaptive k-means clustering algorithm for MR breast image segmentation. *Neural Computing and Applications*, 24(7-8), 1917-1928.

Morris, E. A. & Liberman, L. (2005) *Breast MRI diagnosis and intervention*Springer Science & Business Media.

Otsu, N. (1979) A threshold selection method from gray-level histograms. *IEEE Transactions on Systems, Man, and Cybernetics*, 9(1), 62-66.

Pearson, K. (1895) X. Contributions to the mathematical theory of evolution.—II. Skew variation in homogeneous material. *Philosophical Transactions of the Royal Society of London.(A.)*(186), 343-414.

Pearson, K. (1905) "Das Fehlergesetz und Seine Verallgemeiner-ungen Durch Fechner und Pearson." A Rejoinder. *Biometrika*, 4(1-2), 169-212.

Rasti, R., Teshnehlab, M. & Phung, S. L. (2017) Breast cancer diagnosis in DCE-MRI using mixture ensemble of convolutional neural networks. *Pattern Recognition*, 72, 381-390.

Rivest-Hénault, D., Dowson, N., Greer, P. B., Fripp, J. & Dowling, J. A. (2015) Robust inverse-consistent affine CT–MR registration in MRI-assisted and MRI-alone prostate radiation therapy. *Medical image analysis*, 23(1), 56-69.

Ronneberger, O., Fischer, P. & Brox, T. (2015) U-net: Convolutional networks for biomedical image segmentation, *International Conference on Medical image computing and computer-assisted intervention*. Springer.

Rosset, A., Spadola, L. & Ratib, O. (2004) OsiriX: an open-source software for navigating in multidimensional DICOM images. *Journal of digital imaging*, 17(3), 205-216.

Seiffert, C., Khoshgoftaar, T. M., Van Hulse, J. & Napolitano, A. (2010) RUSBoost: A hybrid approach to alleviating class imbalance. *IEEE Transactions on Systems, Man, and Cybernetics-Part A: Systems and Humans*, 40(1), 185-197.

Spuhler, K. D., Ding, J., Liu, C., Sun, J., Serrano‐Sosa, M., Moriarty, M. & Huang, C. (2019) Task‐based assessment of a convolutional neural network for segmenting breast lesions for radiomic analysis. *Magnetic resonance in medicine*, 82(2), 786-795.

Sutton, E. J., Oh, J. H., Dashevsky, B. Z., Veeraraghavan, H., Apte, A. P., Thakur, S. B., Deasy, J. O. & Morris, E. A. (2015) Breast cancer subtype intertumor heterogeneity: MRI‐based features predict results of a genomic assay. *Journal of Magnetic Resonance Imaging*, 42(5), 1398-1406.

Thomassin-Naggara, I., Trop, I., Lalonde, L., David, J., Péloquin, L. & Chopier, J. (2012) Tips and techniques in breast MRI. *Diagnostic and interventional imaging*, 93(11), 828-839.

Van Aalst, W., Twellmann, T., Buurman, H., Gerritsen, F. A. & ter Haar Romeny, B. M. (2008) Computer-Aided Diagnosis in Breast MRI: Do Adjunct Features Derived from T 2-weighted Images Improve Classification of Breast Masses?, *Bildverarbeitung für die Medizin 2008*Springer, 11-15.

Yu, N., Wu, J., Weinstein, S. P., Gaonkar, B., Keller, B. M., Ashraf, A. B., Jiang, Y., Davatzikos, C., Conant, E. F. & Kontos, D. (2015) A superpixel-based framework for automatic tumor segmentation on breast DCE-MRI, *Medical Imaging 2015: Computer-Aided Diagnosis*. International Society for Optics and Photonics.